\documentclass[11pt]{article}
\usepackage{fullpage,latexsym}
\newcommand{\map}[3]{{#1}:{#2}\rightarrow{#3}}
\newcommand{\bra}[1]{\left\langle{#1}\right|}
\newcommand{\ket}[1]{\left|{#1}\right\rangle}
\newcommand{\tuple}[1]{\langle{#1}\rangle}
\renewcommand{\bowtie}[2]{\ket{#1}\!\bra{#2}}
\newcommand{\ceiling}[1]{\left\lceil{#1}\right\rceil}
\newcommand{\p}{\varphi}

\newtheorem{definition}{Definition}

\newtheorem{theorem}[definition]{Theorem}
\newtheorem{corollary}[definition]{Corollary}

\title{An Intuitive Hamiltonian for Quantum Search}

\author{Stephen Fenner\thanks{Supported in part by NSF Grant
CCR-9996310.  Email \tt{fenner@cs.sc.edu}.} \\
Department of Computer Science and Engineering \\
University of South Carolina \\
CSE-TR-2000-1}

\date{April 24, 2000}

\begin{document}

\maketitle

\begin{abstract}
We present new intuition behind Grover's quantum search algorithm by
means of a Hamiltonian.  
Given a black-box Boolean function $\map{f}{\{0,1\}^n}{\{0,1\}}$ such
that $f(w) = 1$ for exactly one $w\in\{0,1\}^n$, Grover
\cite{Grover:search} describes a quantum algorithm that finds $w$ in
$O(2^{n/2})$ time.  Farhi \& Gutmann \cite{FG:Hamiltonian} show that
$w$ can also be found in the same amount time by letting the quantum
system evolve according to a simple Hamiltonian depending only on $f$.
Their system evolves along a path far from that taken by Grover's
original algorithm, however.  The current paper presents an equally
simple Hamiltonian matching Grover's algorithm step for step.  The new
Hamiltonian is similar in appearance from that of Farhi \& Gutmann,
but has some important differences, and provides new intuition for
Grover's algorithm itself.  This intuition both contrasts with and
supplements other explanations of Grover's algorithm as a rotation in
two dimensions, and suggests that the Hamiltonian-based approach to
quantum algorithms can provide a useful heuristic for discovering new
quantum algorithms.
\end{abstract}

\section{Introduction}

Quantum algorithms can, in theory at least, solve useful problems
faster than classical algorithms.  Two primary families of quantum
algorithms in this regard are algorithms for factoring and discrete
log \cite{Shor:factoring}, and Grover's search
algorithms with quadratic speed-up
\cite{Grover:search,Grover:framework}.

There are many variations on Grover's original algorithm---counting,
starting with partial data, multiple targets, et cetera.  The
algorithm is also surprisingly robust; although the original algorithm
uses the Walsh-Hadamard transform, essentially any unitary operator
will do just as well \cite{}.  Starting with a simple condition on
what transform is used, we will show how Grover's algorithm arises
from a particularly simple---almost naive---intuition about quantum
algorithms.  Our ideas also generalize to variants of Grover's
algorithm.

There have been good explanations in the literature
\cite{Grover:framework,Jozsa:Grover} of how and why fast quantum
search works: the initial state is slowly rotated (in two complex
dimensions) into the target state by repeatedly applying a special
operator known as Grover's iterate.  Using a more physics-based
approach, Farhi \& Gutmann \cite{FG:Hamiltonian} describe an
``analog'' version of quantum search by means of a simple,
time-independent Hamiltonian which transforms any initial state
$\ket{\sigma}$ into some prespecified target state $\ket{w}$ in
optimal time, provided that $\ket{\sigma}$ and $\ket{w}$ are not
orthogonal.  Their analog algorithm rotates the initial state into the
target state in the same time that Grover's ``digital'' algorithm
does, yet their rotation strays far from the intermediate states
reached in the original algorithm by applying Grover's iterate.

We show here that a simple, time-independent Hamiltonian for a system
of qubits results in time evolution matching Grover's iterate exactly.
This Hamiltonian also provides a nice, simple insight into the
workings of the algorithm that is new, to the best of our knowledge.

Our Hamiltonian bears some resemblance to that of Farhi \& Gutmann,
although ours was conceived independently.  Ours differs from theirs
in important respects, however, and may not be as plausible
physically, but it does closely coincide with the iterations of
Grover's algorithm, and thus gives a much closer simulation of a
digital quantum circuit by an analog process and vice versa.  While
Farhi \& Gutmann's Hamiltonian is appealing from a physical point of
view, ours is appealing from an algorithmic perspective.
The heuristic it gives for Grover's algorithm suggests that other
digital quantum algorithms might be found by first looking at analog
versions.  (More work on analog algorithms has been done recently, see
\cite{FGGS:adiabatic} for example.)

\subsection{Structure of the Paper}

We give mathematical preliminaries in Section~\ref{sec:prelims},
including a brief description of Grover's algorithm as described in
\cite{Grover:framework}.  In Section~\ref{sec:look} we show how
Grover's algorithm arises from a simple-minded approach to quantum
search.  The operator we describe there corresponds directly to our
Hamiltonian, which in Section~\ref{sec:Hamiltonians} we compare with
that of Farhi and Gutmann \cite{FG:Hamiltonian}, and show how it
generates Grover's iterate.  Most of our work was done independently
of \cite{FG:Hamiltonian} before it came to our attention, so our
approach to the problem is different.
In Section~\ref{sec:open}, we suggest a Hamiltonian-based approach to
quantum algorithms in general, and we present open problems.

\section{Preliminaries}
\label{sec:prelims}

We work with linear operators over a fixed $N$-dimensional Hilbert
space.  A standard norm on operators is defined as
\[ |A| = \sup_{|v|=1} |Av|, \]
where $A$ is an operator and $|\cdot|$ on the right hand side is the
standard hermitian norm on the Hilbert space.  This norm on operators
satisfies $|AB| \leq |A|\cdot |B|$.  Clearly, all unitary operators
have unit norm.  The exponential map on operators is defined as
\begin{equation}\label{expdef}
e^A = I + A + \frac{A^2}{2!} + \frac{A^3}{3!} + \cdots,
\end{equation}
where $I$ is the identity operator.

If $A$ is skew hermitian ($A^{\dagger} = -A$), then $e^A$ is
unitary.  Conversely, for any unitary $U$ there is a
skew hermitian $A$ such that $U = e^A$.  As in the case with the
exponential function on scalars, we also have
\begin{equation}\label{expapprox}
e^A = \lim_{k\rightarrow\infty} \left(I + \frac{A}{k}\right)^k.
\end{equation}

For an $n$-qubit system we assume the standard basis of states
$\ket{i}$ indexed by classical bit configurations $i \in \{0,1\}^n$.
We use lower-case Roman letters to label basis states, and lower-case
Greek letters to label other (arbitrary) states in the Hilbert space.

\subsection{Grover's Search Algorithm}
\label{sec:Grover}

Here we briefly review Grover's search algorithm.  Fix an integer $n$
and let $N = 2^n$.  Let $f$ be a Boolean-valued function on $n$-bit
strings such that $f(w) = 1$ for exactly one $w$ (the {\em target}),
$0 \leq w < N$ (identifying strings with integers).  A simple version
of Grover's algorithm is to find $w$ via a quantum algorithm where
inputs to $f$ are stored in $n$ qubits, and $f$ is available as a
black box function (oracle) that can be queried by the algorithm.
Alternately, we may assume that $f(y)$ is efficiently computable given
$y$, and embed the computation of $f$ into the quantum circuit.

In this setting, Grover's algorithm (as described in
\cite{Grover:framework,Grover:flexible-search} or \cite{Gruska:quantum})
uses three $n$-qubit unitary transforms:
\begin{enumerate}
\item
an arbitrary, easy-to-compute $U$ such that $\tuple{w|U|0} \neq 0$,
\item
the selected inverter $I_0 = \sum_{0\leq i < N} (-1)^{i=0}
\bowtie{i}{i} = I - 2\bowtie{0}{0}$, and
\item
the selected inverter $I_w = I - 2\bowtie{w}{w}$.
\end{enumerate}
(Here, the formula $i=0$ in the exponent stands for its numerical
truth value---$1$ for true, $0$ for false.)  These combine to form
{\em Grover's iterate}
\begin{equation}\label{Gdef}
G = -UI_0U^{-1}I_w.
\end{equation}
By adjusting $U$ by an appropriate phase factor, we can assume that
$\tuple{w|U|0} = x$ for some real $x > 0$.  This adjustment leaves $G$
unchanged.

Suppose $f$ is as above with $w$ unique such that $f(w) = 1$.
The algorithm starts in the state $\ket{0}$ (all qubits cleared), then
$U$ is applied to get the state
\begin{equation}\label{psidef}
\ket{\psi} = U\ket{0}.
\end{equation}
Next, $G$ is applied repeatedly to $\ket{\psi}$, approximately
$\ceiling{\frac{\pi}{4x}}$ times.  At this point, the system
will be very close to the state $\ket{w}$, so when we now measure the
qubits we get $w$ with high probability.  Note that
\[ I_w = \sum_i (-1)^{f(i)} \bowtie{i}{i}, \]
so $I_w$ can be simulated easily given access to $f$ alone and some
extra work qubits.

In Grover's original presentation, $U = U^{-1}$ is the Walsh-Hadamard
transform on $n$ qubits, and so
\[ \ket{\psi} = 2^{-n/2} \sum_i \ket{i}, \]
whence, $x = \tuple{w|\psi} = 2^{-n/2}$, which yields the quadratic
speed-up in the search.

\section{Quantum Search Revisited}
\label{sec:look}

The point of this section is to show how one might stumble upon
Grover's algorithm by taking a simplistic, almost naive, approach to
quantum search.  The intuition here is not geometric, as it is with
Jozsa \cite{Jozsa:Grover}; rather, it is purely algorithmic in flavor.

We start with the basic observation that if $A$ is a skew hermitian
operator ($A^{\dagger} = -A$) and $0 < \epsilon << 1$, then $I + \epsilon
A$ approximates $e^{\epsilon A}$, which is unitary.  Therefore, $I
+ \epsilon A$ approximates a plausible step in a quantum computation.
By (\ref{expapprox}), we can approximate the action of $e^A$ on a
state by repeatedly applying $I + \epsilon A$ to the state roughly
$1/\epsilon$ times.  The smaller $\epsilon$ is, the better the
approximation.  (In general, it is not certain that $e^{\epsilon A}$
is renderable by a small quantum circuit; it will be in the present
case, though.)

A simple example is when $A = \bowtie{i}{j} - \bowtie{j}{i}$ for some
$i,j \in \{0,1\}^n$, \ $i \neq j$.  Applying $I + \epsilon A$ to a
state $\ket{\p} = \sum_i \alpha_i\ket{i}$ gives
\[ (I + \epsilon A) \ket{\p} = \ket{\p} + \epsilon \alpha_j \ket{i} -
\epsilon \alpha_i \ket{j}. \]
The operator alters $\ket{\p}$ (viewed as a column vector) by adding
an $\epsilon$ fraction of its $j$th component into its $i$th
component, and in exchange, subtracting an $\epsilon$ fraction of its
$i$th component from its $j$th component.  In a sense, we are moving
probability amplitude from state $\ket{j}$ to state $\ket{i}$.
With arbitrary $A$, this swap may take place between many pairs
of components of $\ket{\p}$ at once.

Suppose we are given $n$, $N$, $f$, and $w$ as in
Section~\ref{sec:Grover}.  We start in the state $\ket{\psi} =
N^{-1/2} \sum_i \ket{i}$, which we would like to transform to the
target state $\ket{w}$.  A promising way to do this, given our
considerations above, is to pile positive probability amplitude onto
$\ket{w}$ while taking it away from all the other states evenly.  The
real skew symmetric operator that does this is
\[ A = \left[ \begin{array}{ccccccc}
0      & \ldots & 0      & -1     & 0      & \ldots & 0      \\
\vdots &        & \vdots & \vdots & \vdots &        & \vdots \\
0      & \ldots & 0      & -1     & 0      & \ldots & 0      \\
1      & \ldots & 1      &  0     & 1      & \ldots & 1      \\
0      & \ldots & 0      & -1     & 0      & \ldots & 0      \\
\vdots &        & \vdots & \vdots & \vdots &        & \vdots \\
0      & \ldots & 0      & -1     & 0      & \ldots & 0
\end{array} \right] \]
expressed in the $\{\ket{i}\}$ basis, where the nonzero entries are
all in the $w$th row and $w$th column.  We see that
\[ I + \epsilon A = \left[ \begin{array}{ccccccc}
1        & \ldots & 0        & -\epsilon     & 0        & \ldots & 0        \\
\vdots   &        & \vdots   & \vdots        & \vdots   &        & \vdots   \\
0        & \ldots & 1        & -\epsilon     & 0        & \ldots & 0        \\
\epsilon & \ldots & \epsilon &  1            & \epsilon & \ldots & \epsilon \\
0        & \ldots & 0        & -\epsilon     & 1        & \ldots & 0        \\
\vdots   &        & \vdots   & \vdots        & \vdots   &        & \vdots   \\
0        & \ldots & 0        & -\epsilon     & 0        & \ldots & 1
\end{array} \right]. \]
The $\epsilon$'s on row $w$ have the effect of giving
probability amplitude to $\ket{w}$ while removing it
from all the other states evenly (the column of $-\epsilon$'s).  The
probability amplitude of $\ket{w}$ gains at the expense of an
$\epsilon$ fraction of all the other probability amplitudes.  From
this it is clear that if we start in state $\ket{\psi}$, where all the
probability amplitudes are equal, and apply $I + \epsilon A$ (for some
small $\epsilon$) the right number of times, eventually the state
$\ket{w}$ will dominate.

We note that, using bracket notation,
\[ A = \sqrt{N} ( \bowtie{w}{\psi} - \bowtie{\psi}{w} ). \]
The operator $i\epsilon A$ acts as a Hamiltonian for the time
evolution of the system from $\ket{\psi}$ to $\ket{w}$.  As we'll see
in the next section, for the right value of $\epsilon$, \ $e^{\epsilon
A}$ is exactly two applications of Grover's iterate.

\section{Hamiltonians}
\label{sec:Hamiltonians}

In this section, we give a Hamiltonian for Grover's algorithm, that
is, an operator $H$ such that $e^{-iHt}$ follows the course of the
algorithm as $t$ increases.  It is clear both by geometric
considerations and by the last section that such an operator must
exist.  $H$ is analogous to a previous Hamiltonian $H'$ for quantum
search found by Farhi \& Gutmann \cite{FG:Hamiltonian} which does not
match Grover's algorithm.  We first briefly describe their results,
then describe our Hamiltonian using their framework.

\subsection{Farhi \& Gutmann's Hamiltonian}
\label{sec:FG}

We are given $n$, $N$, $f$ and $w$ as above.  Farhi \& Gutmann
\cite{FG:Hamiltonian} describe a physical, analog way to do quantum search
by first assuming that a Hamiltonian
\[ H_w = E \bowtie{w}{w} \]
is available that distinguishes the target state $\ket{w}$ from all
others by giving it some positive energy $E$ (the other basis states
have energy $0$).  Let $\ket{\sigma}$ be some arbitrary unit vector in the
Hilbert space (the ``start'' state).  We assume $\ket{\sigma}$ is
easy to prepare, so for example, $\ket{\sigma}$ may be $\ket{\psi}$ of
equation (\ref{psidef}).  The goal is to evolve from $\ket{\sigma}$ into
$\ket{w}$.  To search for the state $\ket{w}$, we are allowed to add
some ``driver'' Hamiltonian $H_D$ to $H_w$, provided that $H_D$ does
not depend on the actual value of $w$ at all.  They choose $H_D =
E\bowtie{\sigma}{\sigma}$, so their Hamiltonian is
\[ H' = H_D + H_w = E\left( \bowtie{\sigma}{\sigma} + \bowtie{w}{w} \right), \]
where $E$ is some arbitrary positive value in units of energy.  If
$\ket{\sigma}$ and $\ket{w}$ are not orthogonal, then we can assume as
before that $\tuple{\sigma|w} = \tuple{w|\sigma} = x$ for some $x > 0$ by
adjusting $\ket{\sigma}$ by an appropriate phase factor.

Applying $e^{-iH't}$ to the start state $\ket{\sigma}$ gives the
time-evolution of the system,\footnote{The evolution of a quantum
system under a time-independent Hamiltonian $H$ is actually
$e^{-iHt/\hbar}$.  We choose units so that $\hbar = 1$, and so $Et$ is
a unitless quantity.} which stays in the two-dimensional subspace
spanned by $\ket{\sigma}$ and $\ket{w}$.  Restricting our attention to this
subspace, it is easy to see that $H'$ has eigenvalues $E(1\pm x)$ with
corresponding eigenvectors
\begin{eqnarray*}
\ket{+'} & = & (2+2x)^{-1/2}(\ket{\sigma} + \ket{w}), \\
\ket{-'} & = & (2-2x)^{-1/2}(\ket{\sigma} - \ket{w}).
\end{eqnarray*}
A straightforward calculation yields
\begin{equation}\label{Hprime}
e^{-iH't}\ket{\sigma} = e^{-iEt}\left[\cos(xEt)\ket{\sigma} -
i\sin(xEt)\ket{w}\right].
\end{equation}
When $t = \pi/(2Ex)$, we have
\[ e^{-iH't}\ket{\sigma} = -ie^{-i\pi/(2x)}\ket{w} \]
as desired.

Farhi \& Gutmann observe that if the unit vector $\ket{\sigma}$ is chosen
at random, then the expected value of $x$ is $N^{-1/2}$, making $t =
O(N^{1/2}/E)$.  For constant $E$, this time is the same order of
magnitude as Grover's algorithm.  They show that their time evolution
is optimal up to an order of magnitude for any $w$-independent driver
Hamiltonian $H_D$, even one that varies with time.

\subsection{Another Hamiltonian}
\label{sec:main}

The time evolution of the system according to $H'$ strays far from the
intermediate steps Grover's algorithm.  There surely is a Hamiltonian,
however, whose time evolution matches the steps of Grover's algorithm
exactly, since each step of Grover's algorithm essentially amounts to
a rotation in a two-dimensional space.  We show that this Hamiltonian
can be described very simply: the operator $i\epsilon A$ mentioned at
the end of Section~\ref{sec:look} is exactly the Hamiltonian in
question, for an appropriate value of $\epsilon$ which we will
calculate.

The fact that Grover's iterate can be rendered by a small quantum
circuit then tells us that our intuition of Section~\ref{sec:look} is
justified: the incremental application of $I + \epsilon A$ indeed
corresponds to a legitimate quantum algorithm.

Given $n$, $N$, $f$, $\ket{w}$, $\ket{\sigma}$, $H_w$ and $H_D$ as above,
with $\tuple{w|\sigma} = x > 0$, we define the Hamiltonian
\[ H = \frac{2i}{E} \left[ H_w, H_D \right] = 2iEx (\bowtie{w}{\sigma} -
\bowtie{\sigma}{w}). \]
The rest of this section is devoted to proving the following
\begin{theorem}\label{thm:main}
Assume the special case where $\ket{\sigma} = \ket{\psi}$ and $E = 1$.
Restricted to the $(\ket{\sigma},\ket{w})$-plane, $e^{-iH}$
approximates Grover's iterate $G$ to within $O(N^{3/2})$ in norm.  In
fact, $e^{-iHt_0}$ exactly matches $G$ where
\begin{equation}\label{time-scale}
t_0 = \frac{\pi - 2\arccos x}{2x\sqrt{1-x^2}}.
\end{equation}
On the whole Hilbert space, $e^{-2iH}$ approximates $G^2$ to within
$O(N^{-3/2})$, and $e^{-2iHt_0} = G^2$.
\end{theorem}

For the moment, we allow $E$ to be any positive value and
$\ket{\sigma}$ an arbitrary unit vector with $0 < \tuple{w|\sigma} = x <
1$.  Restricting our attention to the subspace spanned by
$\ket{\sigma}$ and $\ket{w}$, and letting $\theta = \arccos x$, the
eigenvalues of $H$ are seen to be $\pm \frac{1}{2} E\sin 2\theta$ with
corresponding eigenvectors
\[ \ket{\pm} = \frac{1}{\sqrt{2}\sin\theta}\left(e^{\pm
i\theta}\ket{\sigma} - \ket{w}\right). \]
Setting $\eta = E\sin 2\theta = 2Ex\sin\theta$, a routine calculation
shows that
\begin{eqnarray}\label{Hs}
e^{-iHt}\ket{\sigma} & = & \frac{1}{\sin\theta}\left[\sin(\theta -
\eta t) \ket{\sigma} + \sin(\eta t) \ket{w}\right], \\ \label{Hw}
e^{-iHt}\ket{w} & = & \frac{1}{\sin\theta}\left[-\sin(\eta t)
\ket{\sigma} + \sin(\theta + \eta t) \ket{w}\right].
\end{eqnarray}
If $x$ is small, $\theta$ will be close to
$\pi/2$.  For $t = \theta/\eta = \theta/(E\sin 2\theta) \doteq
\pi/(2Ex)$ we have $e^{-iHt}\ket{\sigma} = \ket{w}$.  That is, the system
finds the target state in roughly the same time as with $H'$.

\bigskip

Comparing (\ref{Hprime}) and (\ref{Hs}), we see that the quantum
system evolves significantly differently under the two Hamiltonians
$H'$ and $H$---by more than just a global phase factor.  We now show
how the latter evolution, run for a short time interval, matches a
single step of Grover's algorithm (one application of $G$).  We now
assume $E = 1$ and $\ket{\sigma} = \ket{\psi} = U\ket{0}$ given by
equation~(\ref{psidef}), with $G$ given by (\ref{Gdef}).  We again set
$x = \tuple{\sigma|w} = \tuple{\psi|w} = \cos\theta > 0$, for some $0
< \theta < \pi/2$.

We can express $G$ in the basis $\ket{\sigma}, \ket{w}$:
\begin{eqnarray*}
G & = & - U(I - 2\bowtie{0}{0})U^{-1}(I - 2\bowtie{w}{w}) \\
  & = & - ( I - 2U\bowtie{0}{0}U^{\dagger} ) ( I - 2\bowtie{w}{w} ) \\
  & = & - ( I - 2\bowtie{\sigma}{\sigma} ) ( I - 2\bowtie{w}{w} ) \\
  & = & - I + 2\bowtie{\sigma}{\sigma} + 2\bowtie{w}{w} - 4x\bowtie{\sigma}{w},
\end{eqnarray*}
whence
\begin{eqnarray*}
G\ket{\sigma} & = & (1 - 4x^2)\ket{\sigma} + 2x\ket{w}, \\
G\ket{w} & = & - 2x\ket{\sigma} + \ket{w}.
\end{eqnarray*}
In view of (\ref{Hs}), we solve the equation
\[ \frac{\sin(\eta t)}{\sin\theta} = 2x = 2\cos\theta \]
for $t$ to get the solution
\begin{equation}\label{t-naught}
t_0 = \frac{\pi - 2\theta}{\eta} = \frac{\pi - 2\theta}{\sin 2\theta} =
\frac{\pi - 2\arccos x}{2x\sqrt{1-x^2}},
\end{equation}
as in (\ref{time-scale}).  It is then easy to check that
$e^{-iHt_0}\ket{\sigma} = G\ket{\sigma}$ and that $e^{-iHt_0}\ket{w} =
G\ket{w}$.

Let $S$ be the subspace spanned by $\ket{\sigma}$ and $\ket{w}$ and let
$S^{\perp}$ be its orthogonal complement.  Let $P$ be the
orthogonal projection onto $S^{\perp}$.  We have just shown that
$e^{-iHt_0} = G$ restricted to $S$.  For $\ket{\alpha} \in
S^{\perp}$, clearly $G\ket{\alpha} = -\ket{\alpha}$, while
$e^{-iHt_0}$ leaves $S^{\perp}$ pointwise fixed.  Thus,
\[ e^{-iHt_0} = G + 2P, \]
and since $GP = PG = -P = -P^2$, we have $e^{-2iHt_0} = G^2$ as
expected.

\paragraph{Remark.}
By adding $\frac{\pi}{t_0}P$ to $H$, we get a slightly more
complicated Hamiltonian $\widetilde{H}$ such that
$e^{-i\widetilde{H}t_0} = G$ on the whole Hilbert space.

\bigskip

Finally, we show how close $t_0$ is to $1$, assuming $x << 1$.
Expanding $t_0$ as a power series in $x$, we get
\[ t_0 = 1 + \frac{2}{3}x^2 + O(x^4), \]
and thus
\begin{eqnarray*}
\left| e^{-iHt_0} - e^{-iH} \right| & = & \left| e^{-iH(2x^2/3 + O(x^4))} - I
\right| \\
& = & \left| -\frac{2}{3}iHx^2 + O(x^4)H \right| \\
& = & \frac{2}{3}x^3\sqrt{1-x^2} + O(x^5) \\
& = & \frac{2}{3}x^3 + O(x^5).
\end{eqnarray*}
The second equation comes from expanding the exponential as a power
series.  The third equation holds because $|H| = |\frac{1}{2}\sin
2\theta| = x\sqrt{1-x^2}$.  When $x = N^{-1/2}$, we see that $e^{-iH}$
comes within $O(N^{-3/2})$ of $G + 2P$ in norm, and thus $e^{-2iH}$
comes within $O(N^{-3/2})$ of $G^2$.

\begin{corollary}
If $t = \frac{\pi}{4} N^{1/2}$ then $e^{-iHt}\ket{\sigma} = \ket{w} +
O(1/N)$.
\end{corollary}


\subsection{Discussion}

Farhi and Gutmann show that the Hamiltonian $H'$ finds the target in
optimal time in the following sense: no other Hamiltonian of the same
form---that is, $H_D + E\bowtie{w}{w}$ where $H_D$ has no special
dependence on $w$---can find $\ket{w}$ any faster, even if $H_D$ is
allowed to depend on time.  Our Hamiltonian is clearly not of this
form, so their lower bounds aren't directly applicable here.  Indeed,
it is only by the lower bounds shown for ``digital'' quantum search
\cite{BBBV:quantum} that we know that our Hamiltonian is optimal to
simulate a small digital quantum circuit.  It is an interesting
question whether one can deduce the same lower bound by more direct
means.

\section{Further Research and Open Problems}
\label{sec:open}

We have seen how Grover's algorithm can be described much more simply
using a Hamiltonian than directly with unitary operators.  We don't
present details them here, but variants of Grover's original algorithm
also admit simple Hamiltonian descriptions.  There may be new, yet
unknown quantum algorithms which are more easily described with
Hamiltonians than with unitary operators, and which may indeed be first
discovered by their Hamiltonians.

There are two principal challenges to fashioning a new quantum
algorithm via a Hamiltonian:
\begin{enumerate}
\item
finding an appropriate, and hopefully intuitive, Hamiltonian for the
problem at hand, and
\item
deciding how (or if) the time evolution governed by such a Hamiltonian
approximates a true (digital) quantum algorithm given by a small
quantum circuit.
\end{enumerate}
In the case of Grover's algorithm considered here, we were fortunate
to achieve both goals.  Grover's algorithm came first, however, so we
knew what to shoot for.  Even so, the intuition provided in
Section~\ref{sec:look} may be useful for constructing new
algorithms, or at least viewing other existing algorithms from a
different angle.

Our results thus point to an important general question: when, given a
Hamiltonian on a system of qubits, can the corresponding time
evolution be simulated (even approximately) by a small quantum
circuit?  Is there an easy criterion, based on the structure of the
Hamiltonian itself?  Such a criterion would provide a new way to
prototype new quantum algorithms via Hamiltonians.

Can Farhi \& Gutmann's original $H'$ be simulated efficiently by a
quantum circuit?

Recently, Farhi, {\it et al\@.\/} \cite{FGGS:adiabatic} show how to
solve certain instances of SAT with slowly time-dependent
Hamiltonians (adiabatic evolution).  Their results provide good
physical intuition.  Is there a corresponding algorithmic intuition?

Can one find an intuitive Hamiltonian for a quantum factoring
algorithm?

\section{Acknowledgments}

I would like to thank Frederic Green and Steven Homer for helpful
discussions.


\end{document}